\documentclass[conference]{IEEEtran}
\IEEEoverridecommandlockouts
\RequirePackage{pdf14}

\IEEEpubid{%
  \makebox[\columnwidth][l]{%
    \begin{minipage}[t]{\columnwidth}\vspace{2pt}\footnotesize
      © 2025 IEEE. Personal use of this material is permitted. Permission from IEEE must be obtained for all other uses, in any current or future media, including reprinting/republishing this material for advertising or promotional purposes, creating new collective works, for resale or redistribution to servers or lists, or reuse of any copyrighted component of this work in other works.
    \end{minipage}%
  }%
  \hspace{\columnsep}%
  \makebox[\columnwidth]{}%
}

\usepackage{tikz}
\usetikzlibrary{arrows.meta, shapes.geometric, positioning, shadows.blur}
\usepackage{adjustbox}
\usepackage{cite}
\usepackage{caption}
\usepackage{amssymb,amsfonts}
\usepackage{algorithm}
\usepackage{quantikz}
\usepackage{algpseudocode} 
\usepackage{hyperref}
 \usepackage{booktabs}
\usepackage{doi}
\usepackage{enumitem}
\usepackage{caption}
\usepackage{graphicx}

\usepackage{tikz}
\usetikzlibrary{shapes.geometric, arrows, positioning}
\usepackage{times}
\usepackage{listings}      
\usepackage{xcolor}        

\DeclareCaptionFormat{myformat}{#3}
\usepackage{textcomp}
\usepackage{balance}

\begin{document}

\title{QUEST: QUantum-Enhanced Shared Transportation\\
}

\author{%
    \IEEEauthorblockN{1\textsuperscript{st} Chinonso Onah}
    \IEEEauthorblockA{%
        \textit{Volkswagen Group, Berliner Ring 2, Wolfsburg, Germany}\\
        \textit{Department of Physics, RWTH Aachen University, Germany}\\
        chinonso.calistus.onah@volkswagen.de}
    \and
    \IEEEauthorblockN{2\textsuperscript{nd} Neel Misciasci}
    \IEEEauthorblockA{%
        \textit{Volkswagen Group, Berliner Ring 2, Wolfsburg, Germany}\\
        \textit{CIT School, Technical University of Munich, Germany}\\
        neel.misciasci@volkswagen.de}
    \and
    \IEEEauthorblockN{3\textsuperscript{rd} Carsten Othmer}
    \IEEEauthorblockA{%
        \textit{Volkswagen Group, Berliner Ring 2, Wolfsburg, Germany}\\
        carsten.othmer@volkswagen.de}
    \and
    \IEEEauthorblockN{4\textsuperscript{th} Kristel Michielsen}
    \IEEEauthorblockA{
        \textit{Department of Physics, RWTH Aachen University, Germany}\\
        \textit{Forschungszentrum Jülich, Germany}\\
        k.michielsen@fz-juelich.de}
    \and
}

\IEEEaftertitletext{\vspace{-2.0\baselineskip}}


\maketitle
\IEEEpubidadjcol
\begin{abstract}
We introduce ``Windbreaking-as-a-Service'' (WaaS) as an innovative approach to shared transportation in which a larger ``windbreaker'' vehicle provides aerodynamic shelter for a ``windsurfer'' vehicle, thereby reducing drag and energy consumption. It is intended as a pioneering solution for multi-vehicle highway platooning. As a computational framework to solve the large-scale matching and assignment problems that arise in WaaS, we present \textbf{QUEST} (Quantum-Enhanced Shared Transportation). Specifically, we formulate the pairing of windbreakers and windsurfers -- subject to timing, speed, and vehicle-class constraints -- as a mixed-integer quadratic problem (MIQP). Focusing on a single-segment prototype, we verify the solution classically via the Hungarian Algorithm, a Gurobi-based solver, and brute-force enumeration of binary vectors. We then encode the problem as a Quadratic Unconstrained Binary Optimization (QUBO) and map it to an Ising Hamiltonian, enabling the use of the Quantum Approximate Optimization Algorithm (QAOA) and other quantum and classical annealing technologies. Our quantum implementation successfully recovers the optimal assignment identified by the classical methods, confirming the soundness of the QUEST pipeline for a controlled prototype. While QAOA and other quantum heuristics do not guarantee a resolution of the fundamental complexity barriers, this study illustrates how the WaaS problem can be systematically translated into a quantum-ready model. It also lays the groundwork for addressing multi-segment scenarios and potentially leveraging quantum advantage for large-scale shared-transportation instances.
\end{abstract}


\section{Introduction}
\begin{figure}
    \centering
    \includegraphics[width=1.0\linewidth]{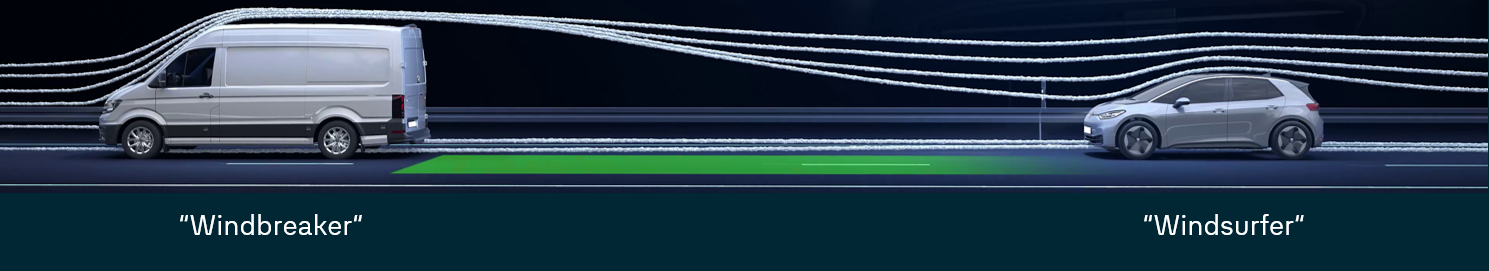}
    \caption{Windbreaking in action.}
    \label{fig: windbreaking}
\end{figure}

The idea of ``Windbreaking-as-a-Service" (WaaS) is to offer a low-tech platooning solution for everyone. On a dedicated platform, not unlike car-pooling, potential donors of wind shadow (``windbreakers") offer spots in their aerodynamic wake, such that other vehicles (``windsurfers") can occupy those spots and benefit from reduced aerodynamic drag and hence a reduced power consumption\cite{duoba2023empirical}. At this stage, we restrict ourselves to {\em pairs} of vehicles: on each portion of the trip, one windbreaker leads exactly one windsurfer (Fig. \ref{fig: windbreaking}). 

\begin{figure}
    \centering
    \includegraphics[width=0.35\linewidth]{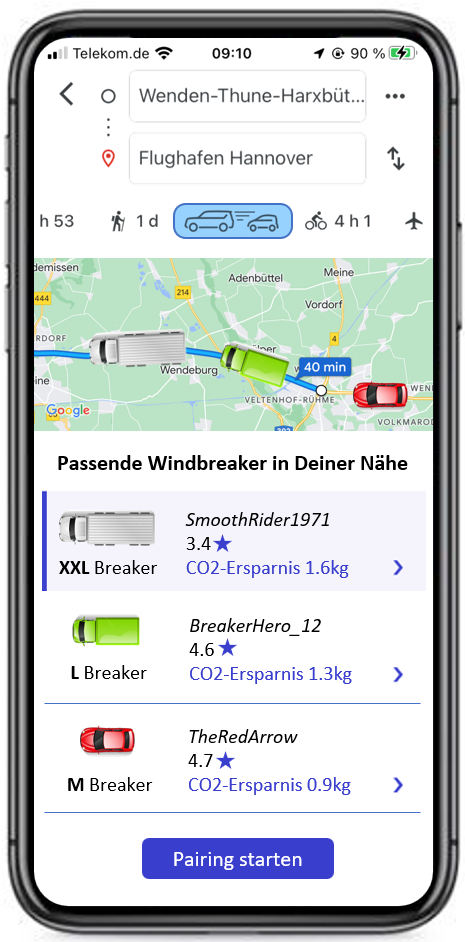}
    \caption{Possible realization of the WaaSApp.}
    \label{fig: waasapp}
\end{figure}

Windsurfers enter start, destination, departure time interval, their own vehicle class (sedan, SUV, van, \ldots) and other preferences (e.\,g.\ their preferred travel speed) into the service app, and a matching algorithm generates a list of suitable windbreakers as indicated in Fig. \ref{fig: waasapp} (or suites of windbreakers, if necessary, for subsequent portions of the highway). The windsurfer chooses one (or a suite) and initiates the pairing. The app issues commands for the rendezvous, and both parties travel jointly up to the destination (or up to the handover point for the next windbreaker in the suite). The jointly driven portions of the trip are on the motorway: pick-up, drop-off and handover locations can be assumed to be at or close to motorway entrances/exits, and handovers should ideally happen on-the-fly, i.\,e.\ not require a stopping of windbreaker or windsurfer.

The matching and assignment needed for WaaS can be viewed as a large-scale combinatorial optimization problem with a rich set of physical and temporal constraints. Specifically, a full-scale WaaS instance may include multiple highway segments and a significant number of surfers and breakers. Timing feasibility constraints may also prevent assignments if the arrival times of a surfer and a breaker at a target road segment differ beyond a threshold. Other side constraints arise in the problem as we lay out in Sec.~\ref{sec:problem_statement}  for the general problem and Sec.~\ref{sec:single-segment-matching} for the single road segment prototype.

Such multi-layered, dynamic assignment structures are strongly believed to be NP-hard in general, since even simpler variants of multi-segment bipartite matchings with side constraints have been shown to inherit complexity from known NP-hard or NP-complete routing and scheduling problems. For instance, Ref.~\cite{garey1979computers} provides a comprehensive treatment of NP‐hardness, including numerous routing and scheduling problems that, even in simpler forms, are NP‐hard. Ref.~\cite{burkard2009assignment} discusses various multi-dimensional assignment problems—which generalize bipartite matchings—and demonstrate that these variants, especially when augmented with additional constraints, are NP‐hard. Even restricted spanning tree problems are NP-hard when required to yield valid routes as solutions\cite{Papadimitriou1982}. Additionally, literature on the \emph{Generalized Assignment Problem (GAP)} shows that once the basic assignment problem is extended with additional resource or side constraints, the problem becomes NP‐hard~\cite{martello1998knapsack}.

Specialized algorithms (e.g., branch-and-cut~\cite{padberg1991branch}, dynamic programming for narrow subproblems~\cite{held1962dp}, or domain-specific heuristics~\cite{glover1989tabu}) might fare well on particular instances, but worst-case scalability remains a looming issue. This reality motivates the search for \emph{alternative paradigms} to combinatorial optimization. Quantum annealing methods~\cite{kadowaki1998quantum}  and \emph{hybrid quantum-classical} algorithms~\cite{farhi2014quantum}, are just a few examples of such paradigms.


\section{Background and Prior Works}\label{sec:priorworks}

Quantum computing has garnered considerable attention due to its potential to solve complex computational problems that classical computers find intractable. For a comprehensive and foundational introduction to quantum computing principles, readers may consult the seminal work by Nielsen and Chuang~\cite{nielsen2000quantum}. Essential quantum algorithms, forming the basis of many contemporary applications, are thoroughly reviewed by Portugal~\cite{portugal2024basicquantumalgorithms}, while practical aspects of programming quantum computers are clearly outlined in the lecture notes by Willsch et al.~\cite{willsch2022lecturenotesprogrammingquantum}.  In particular, the Quantum Approximate Optimization Algorithm (QAOA) has emerged as a promising quantum heuristic for combinatorial optimization problems\cite{farhi2014quantum, Hadfield_2019}. Sturm~\cite{sturm2023theoryimplementationquantumapproximate} offers an accessible yet comprehensive introduction to the theory and practical implementation of QAOA, including instructive case studies executed on IBM quantum hardware. Dalzell et al.~\cite{dalzell2023quantumalgorithmssurveyapplications} provide a detailed survey covering diverse applications and the end-to-end complexity considerations, highlighting the breadth and depth of quantum algorithmic applications while the process of encoding classical combinatorial optimization problems onto quantum architectures is extensively discussed in earlier foundational tutorials by Glover et al.~\cite{glover2018tutorial} and Lucas~\cite{lucas2014ising}.

\subsection{Shared Transportation Problem}

Quantum optimization methods have notable implications for transportation and logistics, where many problems are NP-hard and computationally demanding for classical methods. By mapping tasks such as vehicle routing, traffic flow control, or scheduling onto a Quadratic Unconstrained Binary Optimization (QUBO) form, quantum annealers and gate-model algorithms can in principle provide high quality solutions to these problems even in regimes where classical solvers and heuristics might struggle. Early demonstrations include mapping real-time traffic assignment onto D-Wave hardware~\cite{neukart2017traffic}, extending to small-scale ride-sharing and vehicle-pooling formulations~\cite{cattelan2019combining, cattelan2024modeling} while  recent works have refined QUBO encodings to reduce overhead and incorporate practical constraints~\cite{feld2019hybrid, palackal2023quantumassisted,xie2024feasibility, kadowaki1998quantum}. 

Neukart et al.~\cite{neukart2017traffic} initially demonstrated the applicability of quantum-inspired optimization techniques in traffic flow management. Subsequent studies by Bentley et al.~\cite{bentley2022quantumcomputingtransportoptimization} further investigated quantum computing solutions specifically tailored for transport optimization, exploring routing and scheduling challenges commonly faced in logistics. Complementing this work, Sales and Araos~\cite{sales2023adiabaticquantumcomputinglogistic} examined adiabatic quantum computing approaches, proposing efficient methods for solving logistic transport optimization problems. In a recent review, Zhuang et al.~\cite{zhuang2024quantumcomputingintelligenttransportation} provided a thorough survey of quantum computing applications within intelligent transportation systems. Their review emphasizes emerging quantum solutions, algorithmic developments, and outlines potential future research directions for transportation systems optimization. 

\subsection{Quantum Algorithm}
On the Quantum Algorithm, QAOA~\cite{farhi2014quantum, Hadfield_2019} is a variational quantum algorithm using alternating phase-separation operators (derived from the problem Hamiltonian) and mixing operators to explore the Hilbert space for (near)optimal solutions to optimization problems. Although limited by current circuit depths and noise, QAOA has shown promise on small combinatorial tasks~\cite{harwood2021formulating,qaoabenchmark2021}. In the face of hardware and algorithmic challenges, many near-term applications of quantum computing in logistics adopt hybrid paradigms, where quantum subroutines address particularly challenging subtasks, and classical heuristics handle the rest~\cite{zhou2020quantum, date2019tspqa, feld2019hybrid, palackal2023quantumassisted, venturelli2015job}, demonstrating incremental improvements over purely classical or purely quantum strategies. This has enabled targeted studies to report modest improvements in solution quality or runtime for certain artificially constructed instances or tightly structured embeddings~\cite{guerreschi2019qaoa,dwavebenchmark2020}. Ongoing hardware progress~\cite{gambetta2020ibmroadmap} and more advanced hybrid methods may eventually enable quantum-based solvers to outperform classical algorithms on real-world combinatorial tasks.

\subsection{Motivation}

Field and track studies consistently show that a passenger car or light-duty van travelling in the wake of a taller lead vehicle can cut its traction energy use by
\(\mathbf{10\text{–}13\,\%}\)\cite{duoba2023empirical,NACFE2018,Kim2025}.
If only one vehicle out of ten took advantage of that benefit for
just one-third of its motorway mileage, the resulting US-wide
saving would still approach \(1.3\text{–}1.6\,\text{Mt CO}_2\) per year—
roughly the annual footprint of a mid-sized city of
about \(2\times10^{5}\) inhabitants\cite{UBA2024}.  

Classic ride-sharing assignment without tight timing coupling can be reformulated so that its state space grows merely polynomially and remains tractable for branch-and-cut methods \cite{ma2020rideshare}. However, because the gain is pair-specific and highly dynamic, Windbreaking-as-a-Service leads to timing-window–constrained assignments whose decision space grows combinatorially as
\(2^{\,n^{2}}\)\,\cite{burkard2009assignment}.

Exact mixed-integer solvers for closely related dial-a-ride formulations stall at about \(n\!\approx\!10\) leader–follower pairs \cite{cordeau2003dial}.  By contrast, empirical studies of the Quantum Approximate Optimisation Algorithm show that the circuit depth required for near-optimal solutions rises only logarithmically with problem size \cite{montanezbarrera2024transfer}.   The WaaS matching task therefore remains \emph{exponentially} hard for classical enumeration yet shallow enough in QAOA depth to serve as a stringent benchmark for near-term quantum optimizers.

\subsection{Our Contributions}

This work extends the hybrid quantum-classical optimization paradigm to a novel matching and assignment problem for shared transportation. It differs from prior works on platooning and coalition formation\cite{Zhang2024,Tian2020, Ning2023, Mousavi2019}.


Our study targets a deliberately \emph{low-tech} pilot in which exactly two vehicles — one windbreaker and one windsurfer — coordinate via an off-the-shelf app-like interface (Figure \ref{fig: waasapp}).  This setting avoids the need for vehicle-to-vehicle (V2V) data bus ~\cite{Shladover2015CACC}, SAE Level-4 automation ~\cite{SAEJ3016}, and new platooning legislation. Section ~\ref{sec:problem_statement} outlines the design choices explicitly, and all models in this paper
adhere to the 1-to-1 scope.

Our aerodynamic considerations are based on Ref.~\cite{duoba2023empirical,NACFE2018,Kim2025}, allowing us to focus on the task of formulating complex timing, speed, and vehicle-class constraints into a QUBO problem that can be tackled by quantum annealers or QAOA. This formulation benefits from prior research on penalty-based constraint encoding~\cite{feld2019hybrid, palackal2023quantumassisted, cattelan2019combining}, while the hybrid pipeline design echoes successful approaches in job shop scheduling~\cite{venturelli2015job}.

Additionally, this work aligns with the broader trajectory of applying quantum heuristics to multi-vehicle assignment, where preliminary indications of partial quantum advantage have been reported under carefully selected conditions~\cite{dwavebenchmark2020}. Empirical benchmarks indicate that the proposed approach can achieve solution quality comparable to their classical counterparts while the hardware runtime continues to scale more favourably, reinforcing the potential for quantum hybrid methods to play a meaningful role in next-generation transportation logistics.

\begin{figure}[!htb]
\centering
\begin{tikzpicture}[node distance=1cm, scale=0.74, transform shape]
\tikzstyle{startstop} = [rectangle, rounded corners, minimum width=2.2cm, minimum height=1.2cm, text centered, draw=black, thick, fill=red!15, double]
\tikzstyle{process} = [rectangle, minimum width=3cm, minimum height=1.2cm, text centered, draw=black, thick, fill=gray!10, double, text width=6.5cm, align=center]
\tikzstyle{arrow} = [thick,->,>=latex, shorten >=3pt, shorten <=3pt]

\node (start) [startstop] {Start};

\node (inputs) [process, below=of start] {
\textbf{Inputs}\\
-- Surfers and breakers provide:\\
vehicle classes, timing windows, routes, current location, destination\\
};

\node (ilp) [process, below=of inputs] {
\textbf{Step 1: Bipartite Matching}\\
1. Construct bipartite graph (surfers vs. breakers)\\
2. Define aerodynamic costs as edge weights\\
3. Impose time and velocity constraints
};

\node (solve) [process, below=of ilp] {
\textbf{Step 2: Solve Classically (Validation)}\\
1. Hungarian Algorithm for min-cost assignment\\
2. Brute force (if problem size is small)
};

\node (qubo) [process, below=of solve] {
\textbf{Step 3: Construct QUBO Model}\\
1. Binary variable $x_{s,b}$ for each pairing\\
2. Combine into $x^T Q x$ \\
3. Brute force if problem size is small
};

\node (ising) [process, below=of qubo] {
\textbf{Step 4: QUBO to Ising Hamiltonian}\\
1. $x_i = \frac{1}{2}(1 - Z_i)$ for each variable\\
2. Derive $H_P$ whose ground states are valid solutions
};

\node (qaoa) [process, below=of ising] {
\textbf{Step 5: Apply QAOA}\\
1. Choose depth $p$, initialize $(\gamma, \beta)$\\
2. Uniform superposition of all bitstrings\\
3. Alternate evolving under $U_P$ and a mixing Unitary $U_M$\\
4. Measure final states, extract best solution
};

\node (compare) [process, below=of qaoa] {
\textbf{Step 6: Compare QAOA Solution}\\
1. Interpret measured bitstring as assignment\\
2. Check correctness against classical references
};

\node (outputs) [process, below=of compare] {
\textbf{Outputs}\\
-- Feasible surfer-breaker matchings\\
-- Timing and velocity checks
};

\node (end) [startstop, below=of outputs] {End};

\draw [arrow] (start.south) -- (inputs.north);
\draw [arrow] (inputs.south) -- (ilp.north);
\draw [arrow] (ilp.south) -- (solve.north);
\draw [arrow] (solve.south) -- (qubo.north);
\draw [arrow] (qubo.south) -- (ising.north);
\draw [arrow] (ising.south) -- (qaoa.north);
\draw [arrow] (qaoa.south) -- (compare.north);
\draw [arrow] (compare.south) -- (outputs.north);
\draw [arrow] (outputs.south) -- (end.north);

\end{tikzpicture}
\caption{Flowchart for QUEST Workflow used in the computational parts of this work. It is specialized to a Single-Segment and QAOA. The general pipeline is described in Figure~\ref{fig:quest-pipeline}.}
\label{fig:quest-flowchart}
\end{figure}
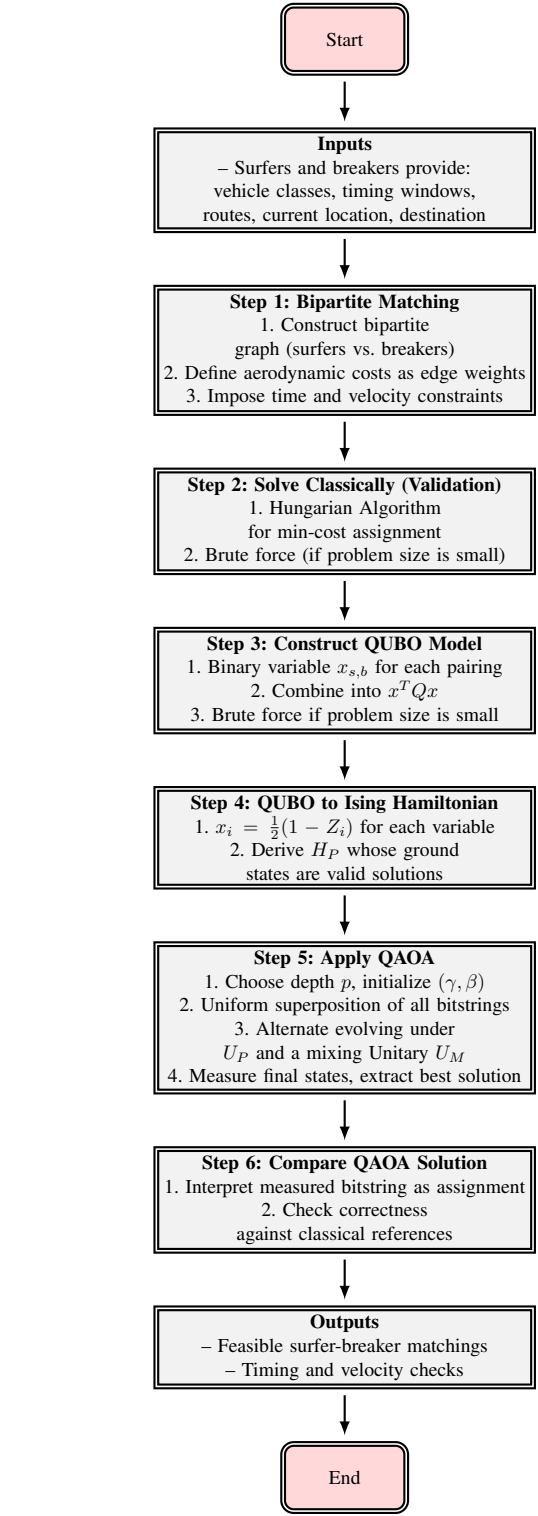

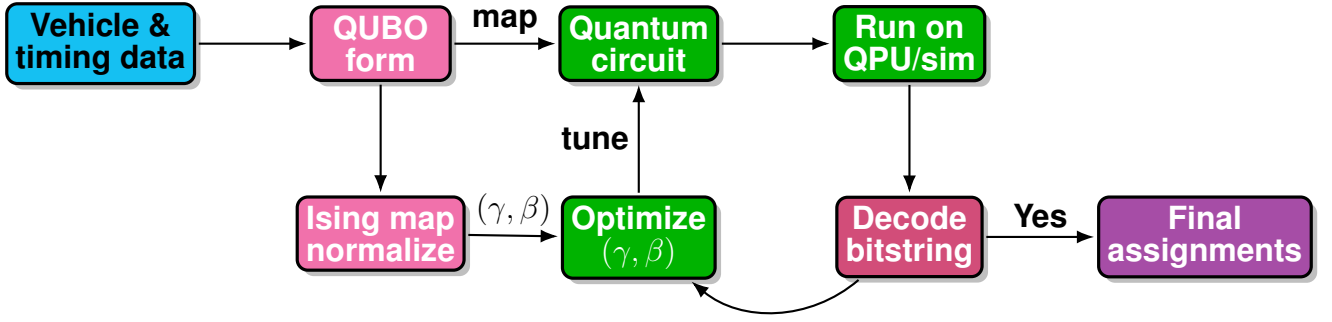
\begin{figure*}[t]
  \centering
  \noindent
  \begin{adjustbox}{%
      max width=\dimexpr2\columnwidth+\columnsep\relax,
      keepaspectratio
    }
    \begin{tikzpicture}[
      scale=1,
      transform shape,
      node distance = 1.5cm and 1.5cm,
      very thick,
      font=\sffamily\bfseries\fontsize{12pt}{12pt}\selectfont,
      >=Latex,
      input/.style   = {draw, fill=cyan!70!white,
                        text=black, align=center,
                        minimum width=1.8cm, minimum height=0.9cm,
                        rounded corners, drop shadow},
      classical/.style = {draw, fill=purple!70!white,
                          text=white, align=center,
                          minimum width=1.8cm, minimum height=0.9cm,
                          rounded corners, drop shadow},
      qubo/.style    = {draw, fill=magenta!70!white,
                        text=white, align=center,
                        minimum width=1.84cm, minimum height=0.9cm,
                        rounded corners, drop shadow},
      quantum/.style = {draw, fill=green!70!black,
                        text=white, align=center,
                        minimum width=1.84cm, minimum height=0.9cm,
                        rounded corners, drop shadow},
      output/.style  = {draw, fill=violet!70!white,
                        text=white, align=center,
                        minimum width=1.8cm, minimum height=0.9cm,
                        rounded corners, drop shadow},
      arr/.style     = {thick, ->, shorten >=1pt, shorten <=1pt}
    ]

    \node[input]     (inputs)  {Vehicle  \&\\ timing data};
    \node[qubo]      (qubo)    [right=of inputs]             {QUBO \\ form};
    \node[qubo]      (ising)   [below=of qubo]               {Ising map\\ normalize};
    \node[quantum]   (qaoa)    [right=1.41cm of qubo]         {Quantum \\ circuit};
    \node[quantum]   (optim)   [below=of qaoa]               {Optimize \\ $(\gamma,\beta)$};
    \node[quantum]   (backend) [right=of qaoa]               {Run on \\ QPU/sim};
    \node[classical] (decode)  [below=of backend]            {Decode \\ bitstring};
    \node[output]    (out)     [right=of decode]             {Final \\ assignments};

    \draw[arr] (inputs)     -- (qubo);
    \draw[arr] (qubo)       -- (qaoa)    node[midway,above,sloped]{map};
    \draw[arr] (qubo)       -- (ising);
    \draw[arr] (ising)      -- (optim)   node[midway,above,sloped]{$(\gamma,\beta)$};
    \draw[arr] (optim)      -- (qaoa)    node[midway,left]{tune};
    \draw[arr] (qaoa)       -- (backend);
    \draw[arr] (backend)    -- (decode);
    \draw[arr,bend left=40,looseness=1.0] (decode) to (optim);
    \draw[arr] (decode)     -- (out)     node[midway,above,sloped]{Yes};

    \end{tikzpicture}
  \end{adjustbox}
  \caption[A general QUEST workflow beyond the single case]{\textbf{A general QUEST workflow beyond the single case.}  
  \textcolor{cyan!80!black}{\textbf{Cyan}} boxes mark the \emph{raw-input layer}: vehicle classes, preferred speeds and departure windows are gathered from both \emph{windsurfers} and \emph{windbreakers}.  
  \textcolor{magenta!80!white}{\textbf{Magenta}} boxes form the \emph{modelling layer}.  First, the data are cast into a quadratic-unconstrained binary optimisation (QUBO) cost function; then the QUBO is transformed into an Ising Hamiltonian and rescaled so that all coefficients lie in \([-1,1]\), making it hardware-friendly.  
  \textcolor{green!90!black}{\textbf{Bright-green}} boxes constitute the \emph{quantum layer}.  The normalised Hamiltonian is compiled into a QAOA circuit, repeatedly executed on a quantum backend while a classical optimiser tunes the angles \((\gamma,\beta)\).  
  \textcolor{purple!85!white}{\textbf{Purple}} boxes show the \emph{classical post-processing layer}.  Measured bitstrings are decoded into surfer-to-breaker matchings; the best result is fed back to the angle optimiser (curved arrow) until convergence.  
  Finally, the \textcolor{violet!80!white}{\textbf{violet}} box delivers the \emph{schedule layer}: a ready-to-use list of optimal pairings with timing and speed instructions.  The solid arrows depict the mandatory data flow; the single curved arrow highlights the optional feedback loop that refines the quantum search.}
  \label{fig:quest-pipeline}
\end{figure*}

Translating the WaaS problem to a Quadratic Unconstrained Binary Optimization (QUBO) model is already a crucial step, since QUBOs are the canonical gateway to many quantum optimization frameworks~\cite{glover2018tutorial, Lucas_2014}. Though the general MIQP objective and constraint structure in large-scale WaaS might be complex, we demonstrate that all the building blocks (i.e., feasibility constraints, objective terms, and penalty terms) can be encoded systematically into a QUBO. Once the QUBO is specified, one can map it to an Ising Hamiltonian to apply QAOA or other quantum heuristics to search for optimal solution(s). A simple computational workflow for the toy problems on a single road segment is presented in Fig. \ref{fig:quest-flowchart} and a more general workflow that applies in general settings even beyond the WaaS problem presented here is depicted in Fig ~\ref{fig:quest-pipeline}.


\begin{figure}[ht]
  \centering
  \begin{tikzpicture}[
    >=Stealth,
    node/.style={circle, fill=black, inner sep=0pt, minimum size=6pt},
    car/.pic={
      \fill[#1] (-0.4,0) rectangle (0.4,0.2);
      \fill[black] (-0.25,0) circle (0.07);
      \fill[black] ( 0.25,0) circle (0.07);
    }
  ]
    \node[node] (v1) at (-2,  1) {};
    \node[node] (v2) at (-2, -1) {};
    \node[node] (c1) at ( 0,  0) {};
    \node[node] (c2) at ( 2,  0) {};
    \node[node] (v3) at ( 4,  1) {};
    \node[node] (m ) at ( 4, -1) {};

    \pic at ($(v1)+(-0.8,0.3)$) {car=green};
    \node[green,anchor=south west] at ($(v1)+(-0.8,0.3)$) {$s_1$};
    \pic at ($(v1)+(-0.8,-0.3)$) {car=blue};
    \node[blue ,anchor=north west] at ($(v1)+(-0.8,-0.3)$) {$b_1$};

    \pic at ($(v2)+(-0.8,-0.3)$) {car=green};
    \node[green,anchor=north west] at ($(v2)+(-0.8,-0.3)$) {$s_2$};

    \pic at ($(c1)+(0,-0.5)$) {car=blue};
    \node[blue,anchor=north] at ($(c1)+(0,-0.5)$) {$b_2$};

    \draw (v1) -- node[above] {$k_1$} (c1);
    \draw (v2) -- node[below] {$k_2$} (c1);
    \draw (c1) -- node[above] {$k_3$} (c2);
    \draw (c2) -- node[above] {$k_4$} (v3);
    \draw (c2) -- node[below] {$k_5$} (m );
  \end{tikzpicture}
  \caption{An illustrative instance of the shared‐transportation problem with $n=2$ surfers and $n=2$ breakers on a motorway network with 5 segments.}
  \label{fig:multi_data_sample}
\end{figure}
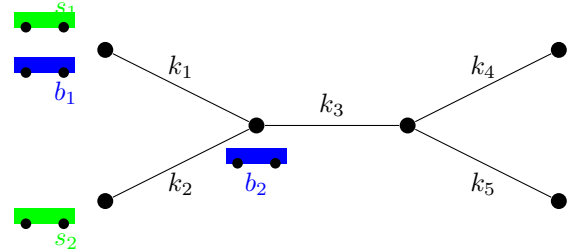




\section{The Matching and Assignment Optimization Problem}\label{sec:problem_statement}

\subsection{Definitions}
\label{definitions}

Given are $B \ge 1$ {\bf windbreakers} and their {\bf fixed} routes. For each windbreaker $b \in \{1, \ldots, B\}$, its route is defined as a series of $N_b$ edges $E_b^1, \ldots, E_b^{N_b}$ of the
motorway graph and their corresponding (expected) starting times $T_b^1, \ldots, T_b^{N_b}$. Every windbreaker $b$ is furthermore described via an average travel speed $V_b$ and a vehicle class $C_b \in \{1,\ldots,5\}$ representing its size from small passenger cars ($C_b=1$) to full-sized trucks ($C_b=5$).

The community of the $S \ge 1$ {\bf windsurfers} is described by a similar set of variables: for each windsurfer $s \in \{1, \ldots, S\}$, there are $n_s$ edges $e_s^1, \ldots, e_s^{n_s}$ that define its route, their corresponding planned starting times $t_s^1, \ldots, t_s^{n_s}$, their {\em preferred} travel speed $v_s$ and their class $c_s \in \{1,\ldots,5\}$. 
While the spatial routes (i.\,e.\ the series of edges) are considered to be fixed, just like the windbreakers' are, the windsurfers' route planning is {\bf variable} in the sense that both the preferred travel speed and the planned starting times for each edge are given as intervals of length $\delta v_s$ and $\delta t_s$, respectively.

We present an optimization model for assigning \emph{windsurfers} to \emph{windbreakers} along a series of consecutive highway segments. 
The objective is to minimize an aerodynamic cost plus handover penalties. 
We then enhance the model with additional constraints to ensure \emph{route connectedness} for each windbreaker and \emph{timing feasibility} for handovers. 
Finally, we illustrate how the model reduces to a simple bipartite matching in the single-segment case.

\subsection{Notation}
Consider a sequence of $K$ highway segments on which a set of $S$ \emph{windsurfers} can be paired with a set of $B$ \emph{windbreakers}. Unlike a single-segment matching, the same surfer or breaker may appear in multiple segments. Thus we rely on \textbf{timing constraints} that prevent a breaker from appearing ``out of nowhere'' on segments it cannot physically reach in time. In what follows, we call this the "No Teleportation Constraint".

We use the following notations:
\begin{itemize}
    \item $S$: number of windsurfers, indexed by $s = 1,\dots,S$.
    \item $B$: number of windbreakers, indexed by $b = 1,\dots,B$.
    \item $K$: number of highway segments, indexed by $k = 1,\dots,K$.
    \item $c_s$ is the vehicle class associated with windsurfer $s$,
    \item $C_b$ is the vehicle class associated with windbreaker $b$,
    \item $d_{s,k}$: distance that \emph{surfer} $s$ must travel to reach segment $k$.
    \item $D_{b,k}$: distance that \emph{breaker} $b$ must travel to reach segment $k$.
    \item $v_s$: speed of surfer $s$.
    \item $V_b$: speed of breaker $b$.
    

    \item $t_{s,k} = \dfrac{d_{s,k}}{v_s}$: arrival time of surfer $s$ on segment $k$.
    \item $T_{b,k} = \dfrac{D_{b,k}}{V_b}$: arrival time of breaker $b$ on segment $k$.

\end{itemize}

\subsection{Constraints}
We define a binary variable $x_{s,b,k} \in \{0,1\}$ for each surfer--breaker pair $(s,b)$ and segment $k$ as follows:
\[
x_{s,b,k} = 1
\quad \Longleftrightarrow 
\]
\[
\text{``surfer $s$ is assigned to breaker $b$ on segment $k$''.}
\]

\paragraph{ Surfer Uniqueness}
Each surfer $s$ must match exactly one breaker on each segment $k$.
\[
\sum_{b=1}^B x_{s,b,k} \;=\; 1,
\quad \forall\,s=1,\dots,S,\; k=1,\dots,K.
\]

\paragraph{Breaker Uniqueness}
Each breaker $b$ can serve exactly one surfer on each segment $k$.
\[
\sum_{s=1}^S x_{s,b,k} \;=\; 1,
\quad \forall\,b=1,\dots,B,\; k=1,\dots,K.
\]

\paragraph{ No Teleportation Constraint}

We do not allow a breaker or surfer to appear ``out of nowhere'' on segment $k$ if they cannot arrive in time to form a valid pairing. 
For instance, if $x_{s,b,k} = 1$, then the arrival times must be close enough to make the assignment physically feasible. 
We therefore impose a time window or maximum allowable gap $\Delta$: 
\[
x_{s,b,k} = 1
\;\Longrightarrow\;
\bigl|\,t_{s,k} - T_{b,k}\bigr| \;\le\;\Delta.
\]
If the surfer and breaker arrive at segment $k$ far apart in time, $x_{s,b,k}$ must be 0 and the pairing is excluded.

\subsection{Cost Function}

\subsubsection{Aerodynamic Cost}
The objective is to minimize overall CO$_2$ emission (i.\,e.\ power consumption). In the considered setting, the overall CO$_2$ emission is controlled by the power consumption of the windsurfers; more precisely, its aerodynamic portion. It depends on {the surfer's vehicle class}, the travel speed of the breaker (which is adopted by the surfer) and the windbreaking efficiency. 
    
We let $w_{s,b,k}$ be the weighted cost of assigning windbreaker $b$ to windsurfer $s$ on segment $k$:
\begin{equation}
w_{s,b,k} 
\;=\;
c_s\, V_b^2\, \bigl[\,1 - f(C_b - c_s)\bigr]\, l_{s,k},
\label{eq:asseqn}
\end{equation}
where:
\begin{itemize}
    \item $f(C_b - c_s)$ is an ``efficiency function'' depending on the class difference $C_b - c_s$, For a given pairing of a windbreaker $b$ and a windsurfer $s$, this efficiency can be assumed to be a linear function $f$ of the vehicle class difference $d:=C_b-c_s \in \{-4,\ldots,4\}$ as follows:
\begin{equation}
    f\left(d\right)=\frac{1}{24}\left(d+4\right)\,,
\end{equation}
\footnote{Replacing the linear law by quadratic dependence (or any powers of the class difference) leaves the optimization problem unchanged; only the numerical entries \(w_{s,b,k}\) will differ.}
such that $f(-4)=0$ (small breaker leads a large surfer), and $f(4)=1/3$ as maximum efficiency for the best possible breaker-surfer combination.
    \item $l_{s,k}$ is the length of segment $k$ relevant to windsurfer $s$.
    \item $V_b$ is the velocity of the breaker. 
\end{itemize}

Then the total \emph{assignment cost} is
\begin{IEEEeqnarray}{rCl}
\sum_{k=1}^K \sum_{s=1}^S \sum_{b=1}^B
w_{s,b,k} \; x_{s,b,k}.
\end{IEEEeqnarray}

\subsubsection{Time-Dependent Handover Cost}

When surfer $s$ travels behind a suitable breaker $b$ on road segment $k$, it is desirable to avoid switching from breaker $b$ in segment $k$ to breaker $b'$ in segment $k'$.  We capture this by introducing a quadratic term $x_{s,b,k}\,x_{s,b',k'} =  1$  when surfer $s$ switches from $b$ to $b'$. This handover requires a certain amount of time to be orchestrated. To capture these possibilities we define the coefficient of this quadratic term as:
\begin{IEEEeqnarray}{rCl}
\Delta T_{b,b',k,k'} 
\;=\; 1 + 
\bigl|\,
T_{b,k}
\;-\;
T_{b',k'}
\bigr|.
\end{IEEEeqnarray}
If the two breakers are present at the beginning of the road segment to initiate a handover, the time difference contribution to the handover objective vanishes. In other words, we look at how far apart (in time) breaker $b$ is from the new breaker $b'$ on the new segment $k'$. 

The total handover objective becomes:
\begin{IEEEeqnarray}{rCl}
\IEEEeqnarraymulticol{3}{l}{
    \sum_{s=1}^S \sum_{k=1}^K \sum_{b=1}^B \sum_{b'=1}^B \sum_{k'=2}^K
    \Delta T_{b,b',k,k'} \, x_{s,b,k}\,x_{s,b',k'}
}
\end{IEEEeqnarray}
Where $k$ and $k'$ are adjacent segments. 
Note that a time-dependent handover cost does \emph{not} make the ``no teleportation'' constraint redundant. The latter imposes $\bigl|T_{s,k} - T_{b,k}\bigr|\le \Delta$ as a strict feasibility condition which forbids the assignment if violated, \emph {regardless} of cost considerations. On the other hand, the "Time-Dependent Handover Cost" is a part of the objective to be optimized for all possibilities not in violation of the teleportation constraint.

\subsubsection{The Optimization Objective}
Combining the segment assignment cost and the handover cost we obtain the following objective:

\begin{IEEEeqnarray}{rCl}
\min_{\{x_{s,b,k}\}} &&
\underbrace{\sum_{k=1}^K \sum_{s=1}^S \sum_{b=1}^B w_{s,b,k}\, x_{s,b,k}}_{\text{Assignment Cost}} \nonumber\\[1mm]
&& {}+ \underbrace{\sum_{s=1}^S \sum_{k=1}^K \sum_{k'=1}^K \sum_{b=1}^B \sum_{b'=1}^B \Delta T_{b,b',k,k'}\, x_{s,b,k}\, x_{s,b',k'}}_{\text{Handover Cost}}
\end{IEEEeqnarray}

Subject to the following constraints:

\begin{IEEEeqnarray}{l}
\textbf{(i) Surfer Uniqueness:} \nonumber\\
\hspace{1.5em} \sum_{b=1}^B x_{s,b,k} = 1,\quad \forall\, s,\,k; \nonumber\\[1mm]
\textbf{(ii) Breaker Capacity:} \nonumber\\
\hspace{1.5em} \sum_{s=1}^S x_{s,b,k} \le 1,\quad \forall\, b,\,k; \nonumber\\[1mm]
\textbf{(iii) Timing Feasibility:} \nonumber\\
\hspace{1.5em} x_{s,b,k} = 1 \Longrightarrow \left|t_{s,k} - T_{b,k}\right| \le \Delta.
\end{IEEEeqnarray}

Where $k$ and $k'$ are adjacent segments.

\section{WaaS on a Single Road Segment}\label{sec:single-segment-matching}

\begin{figure}[htbp]
    \centering
    
    \fbox{\includegraphics[width=0.78\linewidth]{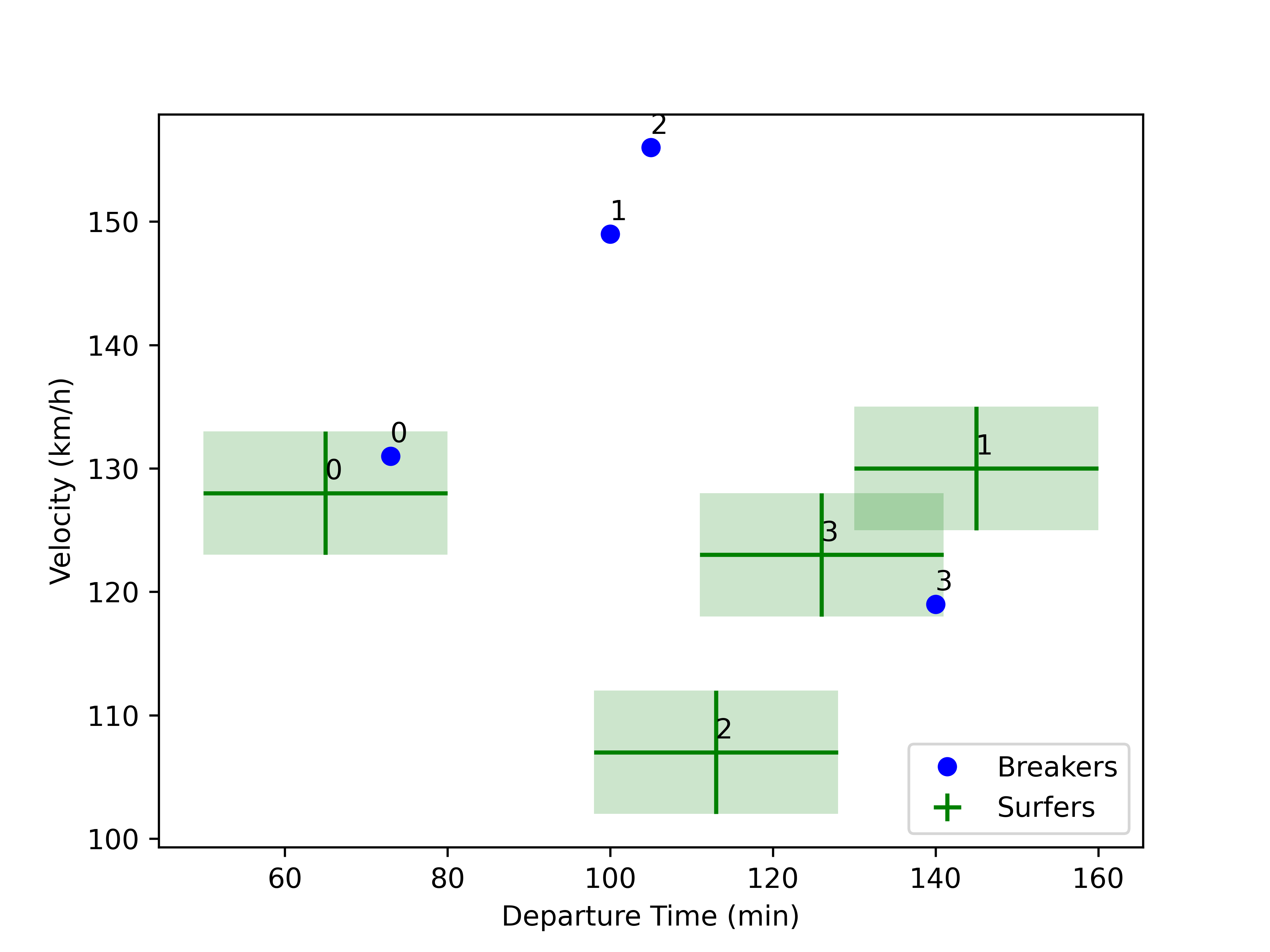}}
    \caption{An illustrative instance of the WaaS assignment problem with $n=4$ surfers and $n=4$ breakers. For a given edge, a matching between a breaker $b$ and surfer $s$ would be regarded as valid, if $V_b \in [v_s-\delta v_s/2,v_s+\delta v_s/2]$ and $T_b \in [t_s-\delta t_s/2,t_s+\delta t_s/2]$ (here: $\delta v_s=10$km/h, $\delta t_s=30$min $\forall s$). The objective is to select a set of edges such that each surfer is assigned to exactly one breaker while minimizing the total cost of all chosen edges.}

    \label{fig: data_sample}
\end{figure}


In the following, we specialize the general problem to the case of a single road segment, i.\,e.\ to $|K|=1$.
Let the set of surfers be \(\mathcal S = \{ 1, \dots, S\}\) and the set of breakers be \(\mathcal B = \{ 1, \dots, B\}\).
We form a bipartite graph \(G(\mathcal S \cup \mathcal B , \mathcal E)\) where each edge \((s,b)\in \mathcal E\) indicates that surfer \(s \in \mathcal S\) can potentially be matched to breaker \(b \in \mathcal B\). For each possible edge \((s,b)\in \mathcal E\), we define a binary variable \(x_{s,b}\in\{0,1\}\) such that:
\[
x_{s,b} = 
\begin{cases}
1, & \text{if surfer $s$ is assigned to breaker $b$,}\\
0 & \text{otherwise.}
\end{cases}
\]

In the quantum treatment, each qubit in the computational basis corresponds to an edge \((s,b)\). Let \(w_{s,b}\) be the weight of that edge, incorporating aerodynamic efficiency and constraint offsets one obtains:
\begin{equation}
w_{s,b}
\;=\;
c_s\, V_b^2\, \bigl[\,1 - f(C_b - c_s)\bigr]
\;+\;
\lambda_1 \,\Delta_{s,b}^{T}
\;+\;
\lambda_2 \,\Delta_{s,b}^{V},
\label{eq:weight_formula}
\end{equation}
where \(\lambda_1, \lambda_2 \ge 0\) are scaling factors, and \(\Delta_{s,b}^{T}\) and \(\Delta_{s,b}^{V}\) capture time and velocity preferences of surfer \(s\) for breaker \(b\) as follows:
\begin{eqnarray}
\Delta_{s,b}^{T} &=&
\begin{cases} 
|t_s-T_b| & \text{if } |t_s-T_b| > \delta t_s/2, \\ 
0 & \text{otherwise,}
\end{cases}
\end{eqnarray}
and similarly for $\Delta_{s,b}^{V}$, with $\delta t_s$ and $\delta v_s$ being surfer-individual flexibility intervals for departure time and preferred travel speed, respectively.

We further assume $B=S$ and require each surfer \(s \in \mathcal S\) to be matched to exactly one breaker, and each breaker \(b \in \mathcal B\) to accept exactly one surfer. That is:
\[
\sum_{b \in \mathcal B} x_{s,b} = 1
\quad \forall s\in \mathcal S,
\qquad
\sum_{s \in \mathcal S} x_{s,b} = 1
\quad \forall b\in \mathcal B.
\]
To transform these into an unconstrained objective, we add squared-penalty terms with a hyperparameter \(\lambda_3\):

\begin{equation}\label{eq:quad_penalties}
\lambda_3 \sum_{s\in \mathcal S} 
\Bigl(1 - \sum_{b\in \mathcal B} x_{s,b}\Bigr)^{2}
\;+\;
\lambda_3 \sum_{b\in \mathcal B}
\Bigl(1 - \sum_{s\in \mathcal S} x_{s,b}\Bigr)^{2}.
\end{equation}

With only one segment under consideration, there are no consecutive‐segment transitions and thus no handover cost. Hence, our \emph{Quadratic Unconstrained Binary Optimization} (QUBO) problem is:

\begin{IEEEeqnarray}{rCl}
\min_{\{x_{s,b}\}} & & \sum_{(s,b)\in \mathcal E} w_{s,b}\,x_{s,b} \nonumber\\[1mm]
& & +\, \lambda_3 \sum_{s\in \mathcal S}\Bigl(1 - \sum_{b\in \mathcal B} x_{s,b}\Bigr)^2 \nonumber\\[1mm]
& & +\, \lambda_3 \sum_{b\in \mathcal B}\Bigl(1 - \sum_{s\in \mathcal S} x_{s,b}\Bigr)^2.
\label{eq:final_qubo}
\end{IEEEeqnarray}

With \(n:=S(=B)\), we can list all pairs \((s,b)\) in a vector:
\[
x \;=\; \bigl(
x_{1,1},\, x_{1,2},\,\dots,\, x_{1,n},\,
x_{2,1},\,\dots,\, x_{n,n}
\bigr)^T.
\]
This vectorized form yields the canonical QUBO equation: 

\begin{equation}
\min \, x^T Q x
\label{eq: qubo}
\end{equation}

\begin{equation}
\text{with} \quad Q_{ee} = w_{s,b} + \lambda_1 \Delta_{s,b}^{T} + \lambda_2 \Delta_{s,b}^{V}- 2 \lambda_3
\end{equation}

\begin{equation}
\text{and} \quad Q_{ee'} = 
\begin{cases} 
\lambda_3 & \text{if } e \text{ and } e' \text{ share a node} \\ 
0 & \text{otherwise.}
\end{cases}
\end{equation}tackled by standard QUBO solvers on classical and quantum hardware.






\section{Classical Solution Methods}
\label{sec:classical-methods}

Classical methods for solving the single-segment windbreaker--windsurfer assignment are grounded in standard bipartite matching techniques. When each surfer can be assigned to exactly one breaker (and vice versa), and when all feasibility constraints for speeds and timings are satisfied (or implemented as soft penalties), the pairing problem reduces to a classical \emph{Bipartite Matching Problem}, which is classically solvable in polynomial time. For the purpose of benchmarking, we implemented the Hungarian (a.k.a.\ Kuhn--Munkres) algorithm\cite{munkres1957algorithms}, solved the resulting  QUBO  problem  via exhaustive search and via Gurobi solver\cite{gurobi}.

\subsection{Hungarian Algorithm}

In our prototype scenario, we construct a cost matrix $\mathbf{C}$ whose rows correspond to windsurfers and columns correspond to windbreakers, and each entry $\mathbf{C}_{(s,b)}$ represents the aerodynamic or energy cost (plus suitably scaled penalty terms) of assigning surfer $s$ to breaker $b$. The Hungarian algorithm then returns optimal row and column indices minimizing the total cost.  We utilized the linear-sum-assignment method from the scipy library\cite{virtanen2020scipy} to realize the algorithm and solve the single-segment matching problem to optimality.

\begin{figure}[htbp]
    \centering
    \fbox{\includegraphics[width=0.79\linewidth]{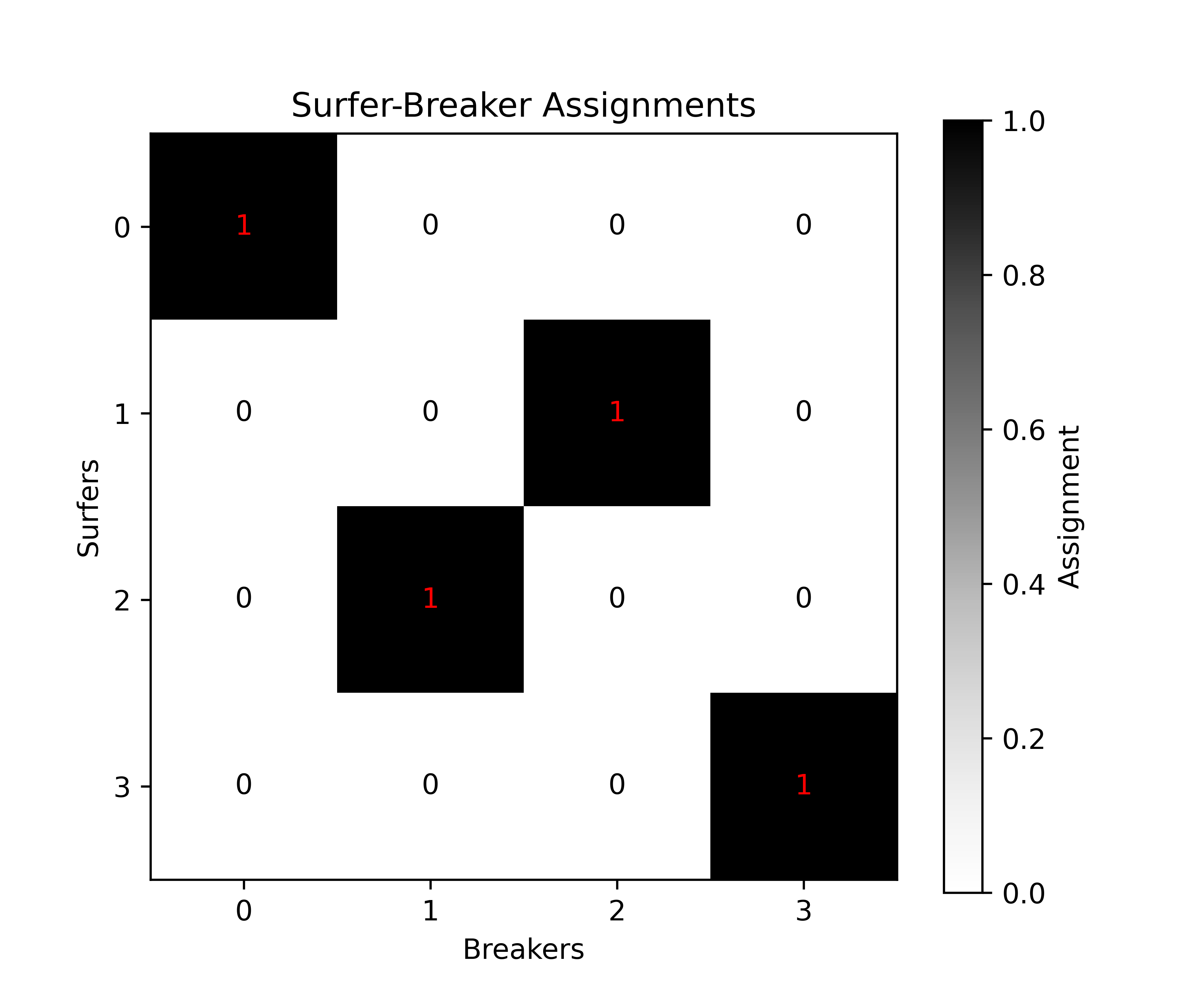}}
    \caption{An example assignment matrix for $n=4$ surfers vs.\ breakers. As expected, the Hungarian method and brute force exhaustive enumeration lead to the same assignments. In this example: Surfer 0 is assigned to Breaker 0, Surfer 1 is assigned to Breaker 2, Surfer 2 is assigned to Breaker 1, Surfer 3 is assigned to Breaker 3. In agreement with visual inspection of Figure \ref{fig: data_sample}.}
    \label{fig:hungarian-demo}
\end{figure}

\subsection{QUBO Formulation and Classical Solvers}

The QUBO matrix $\mathbf{Q}$ in Eq~\ref{eq: qubo} incorporates diagonal terms, reflecting the cost contribution of each surfer--breaker pair, and the off-diagonal penalty terms preventing multiple surfers from matching to the same breaker (and vice versa). Though QUBOs are often associated with quantum annealers or gate-model quantum algorithms (e.g.\ QAOA), they can also be tackled by classical methods. Thus, a classical QUBO solver was implemented with Gurobi. Additionally, QUBO instances were solved by exhaustively enumerating all $2^{n^2}$ binary vectors.  Specifically, we create a vector $\mathbf{x}$ of length $n^2$, where each component $x_{sb}$ represents whether surfer $s$ is matched to breaker $b$ and then compute the QUBO energy $ \mathbf{x}^\top \mathbf{Q}\,\mathbf{x}$ for each vector.  The vector yielding the minimum energy is the global optimum.  This brute force approach confirms correctness of other solvers (e.g.\ Hungarian or QUBO-based heuristics) on small test scenarios,  and serves as a definitive reference for verifying solutions.


\section{Quantum Optimization Method}\label{sec:qaoa}

\subsection{Mapping the QUBO to a Quantum Hamiltonian}

To solve an optimization problem on a quantum computer, its QUBO problem must be mapped to a quantum Hamiltonian~\cite{Lucas_2014}. This is derived in the standard way by expressing a symmetric QUBO matrix in terms of its components and transforming the upper triangular part into an Ising Hamiltonian. Consider the following QUBO matrix:

\begin{equation}
x^T Q x = \sum_{i,j} Q_{ij} x_i x_j
\end{equation}

where \( x_i \in \{0, 1\} \) are binary variables, and \( Q_{ij} \) is the matrix defining the quadratic coefficients. Each binary variable \( x_i \) is mapped to a qubit using the relation:

\begin{equation}
x_i = \frac{1}{2} (I - Z_i)
\end{equation}

where \( Z_i \) is the Pauli-Z operator acting on the \( i \)-th qubit, and \( I \) is the identity operator, effectively transforming the binary variable \( x_i \) into a Pauli-Z operator. See Ref~\cite{Lucas_2014} for details.

The full Hamiltonian becomes

\begin{equation}
H = \sum_{i} \frac{Q_{ii}}{2} (I - Z_i) + \sum_{i < j} \frac{Q_{ij}}{4} \left( I - Z_i - Z_j + Z_i Z_j \right)
\end{equation}

The constant identity terms do not influence the optimization landscape of the QUBO problem and can be omitted.



\subsection{Standard QAOA Protocol}

Since each qubit in the computational basis corresponds to an edge \((s,b)\), $D = |S| \cdot |B| $ is the total number of qubits required to solve the problem posed in Sec.~\ref{sec:single-segment-matching}. To implement the Quantum Approximate Optimization Algorithm\cite{farhi2014quantum}, the first step is to prepare an initial state  $\lvert 0 \rangle^{\otimes D}$. Then, we apply to $\lvert s \rangle$ 
the family of phase-separation operators $U_P(\gamma)$ that depend on the 
problem Hamiltonian $H_P$ as 

\begin{equation}
U_P(\gamma) = e^{-i \gamma H_P.}
\end{equation} 

\begin{figure}
\centering
\resizebox{\columnwidth}{!}{%
\begin{quantikz}[row sep={0.6cm,between origins}, column sep=0.5cm]
\lstick[wires=4]{$\ket{0}^{\otimes n}$}
& \gate[4]{\text{Init State}}
& \gate[4]{U_P(\gamma)}
& \gate[4]{U_M(\beta)}
& & \meter{}\\
& & & & & \meter{}\\
& & & & & \meter{}\\
& & & & & \meter{}
\end{quantikz}%
}
\caption{A Single Layer of QAOA Circuit}
\label{fig: qaoa_layers}
\end{figure}
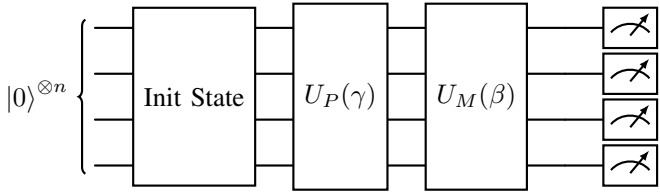

Next, we apply the mixing-operators $U_M(\beta)$ where $U_M(\beta)$ must preserve the initial state $\lvert s \rangle$  and 
provide transitions between all its basis vectors. A QAOA circuit then consists of $p$ alternating layers of $U_P(\gamma)$ and $U_M(\beta)$ applied to a suitable 
initial state $\lvert s \rangle$ as follows (See Fig~\ref{fig: qaoa_layers}):

\begin{equation}
\lvert \gamma, \beta \rangle = e^{-i \beta_p H_M} e^{-i \gamma_p H_P} \ldots e^{-i \beta_1 H_M} e^{-i \gamma_1 H_P} \lvert s \rangle 
\end{equation}

A computational basis measurement over the state $\lvert \gamma, \beta \rangle$ 
returns a candidate solution state $\lvert z \rangle$ close to the ground state of the Ising Hamiltonian with high probability and by the same token, a near optimal solution to the original problem. If we consider the objective function value of the original problem to be $f(z)$, then QAOA succeeds with probability $\lvert \langle z \lvert \gamma, \beta \rangle \rvert^2$ if $\lvert z \rangle$ is an ideal solution. The goal of QAOA 
and its many variants\cite{Hadfield_2019} is to prepare a state $\lvert \gamma, \beta \rangle$, from which we can sample 
a solution $z$ with a high value of $f(z)$.

In the generic QAOA\cite{farhi2014quantum} the initial state, mixer and problem unitaries are given respectively: 

\begin{equation}
| s \rangle = | + \rangle | + \rangle \ldots | + \rangle
\end{equation}

\begin{equation}
U(\beta) = \prod_{j} e^{-i \beta X_j}
\end{equation}

\begin{equation}
U_P(\gamma) = e^{-i \gamma H_P.}
\end{equation} 

Therefore:

\begin{equation}
\langle \psi | U^\dagger(\gamma_p)U^\dagger(\beta_p) H U(\beta_p) U(\gamma_p) | \psi \rangle = f(\beta, \gamma)
\end{equation}

is a valid QAOA protocol for our problem and our $f(z)$ is \begin{equation}
f(z) = f(\beta, \gamma)
\end{equation}

\subsection{Hamiltonian Normalization Scheme}
It could arise that the problems are defined by large coefficients, leading to numerical instability and convergence difficulties. This is addressed through the so called Hamiltonian normalization. Below we outline the main ideas. See  scheme~\cite{montanezbarrera2024universalqaoaprotocolevidence} for details. 

Given a $QAOA$ cost Hamiltonian $H$, it can be generically expressed as:
\begin{equation}
H \;=\; \sum_{k} c_k \, P_k,
\end{equation}
where $P_k$ are Pauli operators (or tensor products of Pauli $I, Z$), and $c_k \in \mathbb{R}$ are the corresponding real coefficients. We define
\begin{equation}
\alpha \;=\; \frac{1}{\max_{k} \lvert c_k \rvert}.
\end{equation}
Our \textit{normalization} scheme consists of scaling the entire Hamiltonian $H$ by $\alpha$. The new (normalized) Hamiltonian is:
\begin{equation}
H_{\text{norm}} 
\;=\; \alpha H 
\;=\; \sum_{k} \alpha \, c_k \, P_k
\;=\; \sum_{k} \left(\frac{c_k}{\max_{k'} \lvert c_{k'} \rvert}\right) P_k.
\end{equation}
Because $\alpha = 1 / \max \lvert c_k\rvert$, the largest coefficient in magnitude of $H_{\text{norm}}$ is exactly $1$.

Multiplying a Hamiltonian by a \textbf{positive} scalar $\alpha$ preserves the set of ground states and their relative ordering. Concretely:

Let $\lvert \psi_{\text{gs}} \rangle$ be a ground state of $H$ with energy $E_{\text{gs}}$, so
\begin{equation}
H \lvert \psi_{\text{gs}} \rangle 
\;=\; E_{\text{gs}} \, \lvert \psi_{\text{gs}} \rangle.
\end{equation}
Then for $H_{\text{norm}} = \alpha H$,
\begin{equation}
\alpha H \lvert \psi_{\text{gs}} \rangle 
\;=\; \alpha E_{\text{gs}} \lvert \psi_{\text{gs}} \rangle.
\end{equation}
Hence $\lvert \psi_{\text{gs}} \rangle$ is also a ground state of $H_{\text{norm}}$, with the ground energy scaled by $\alpha$.

In the classical Ising/QUBO interpretation, if $E(\mathbf{x})$ is the cost (energy) of a configuration $\mathbf{x}$ in the original Hamiltonian, then the new cost is $E'(\mathbf{x}) = \alpha E(\mathbf{x})$. Since $\alpha > 0$, minimizing $E(\mathbf{x})$ is equivalent to minimizing $\alpha E(\mathbf{x})$. This technique helps us to avoid very large coefficients in the Hamiltonian and is very helpful for convergence and numerical stability in QAOA algorithm\cite{montanezbarrera2024universalqaoaprotocolevidence}.

\subsection{QAOA Implementation}
\label{sec:qaoa_implementation}

We implemented the Quantum Approximate Optimization Algorithm (QAOA)~\cite{farhi2014quantum} using the \emph{Qiskit} SDK\cite{qiskit}. As discussed in Sec.~\ref{sec:qaoa}, in order to avoid excessively large coefficients and improve the numerical stability of the QAOA routine, we normalize the Hamiltonian by dividing all coefficients by the maximum absolute coefficient value. With the normalized cost Hamiltonian, we generate a parameterized QAOA \emph{ansatz} that starts from a uniform superposition of all computational basis states. 
A single round (or \emph{layer}) of QAOA consists of one application of \(U_P(\gamma)\) followed by one application of \(U_M(\beta)\) while a higher-depth QAOA repeats these layers to form increasingly expressive quantum circuits. 

\subsection{Classical Parameter Optimization}
The parameters \(\{\gamma, \beta\}\) appearing in the phase-separation and mixing operators are optimized via a classical routine. After each circuit execution, the resulting expectation value of the Hamiltonian is measured, and the classical optimizer adjusts the parameters to minimize that energy. This is iterated until a convergence criterion is reached or a preset iteration limit is exceeded. To determine an optimal set of QAOA parameters, we performed a brute-force grid search over the \(2p\)-dimensional parameter space, where \(p\) is the number of QAOA layers. Each layer contributes a phase-separation parameter \(\gamma\) and a mixing parameter \(\beta\), and we discretized their allowed ranges (set here to \([0,\pi]\)) using a fixed grid density. For every point on this grid, we evaluated the cost function—i.e., the expectation value of the problem Hamiltonian—by executing the corresponding QAOA circuit on a simulator backend with a small number of shots. The parameters that yielded the lowest cost were then selected as optimal.\footnote{A  universal QAOA protocol based on simple linear schedules was recently introduced in \cite{montanezbarrera2024universalqaoaprotocolevidence} to avoid the classical parameter optimization overhead which is known to be NP-Hard\cite{bittel2021training} }

Due to the well-known phenomenon of transfer learning in QAOA~\cite{montanezbarrera2024transfer, zhou2020quantum, TransferabilityOO}, the optimal parameter set obtained from this brute-force search was effective across a range of problem sizes. In our experiments, theA  universal QAOA protocol based on simple linear schedules was recently introduced in \cite{montanezbarrera2024universalqaoaprotocolevidence} to avoid the classical parameter optimization overhead which is known to be NP-Hard\cite{bittel2021training}  same parameter values were found to solve problems from 2 surfer–breaker pairs (resulting in a \(4\)-qubit problem) up to 4 surfer–breaker pairs (corresponding to a 16-qubit problem). This concentration effect implies that the optimal parameters tend to cluster in a narrow region of the parameter space, thereby reducing the complexity of the classical optimization loop in QAOA implementations. We initialized other QAOA runs with these parameters and used different interations of Nelder-Meader classical optimizer implemented in scipy library to solve all problems presented.
Following the optimization of the QAOA parameters, we measure the final state multiple times, obtaining a distribution over bitstrings. The bitstring with the highest sampling frequency often represents a low-energy (i.e., high-quality) solution. In the WaaS context, a bitstring indicates the specific windsurfer--windbreaker assignments that potentially minimize total cost and satisfy timing or capacity constraints. Additional classical post-processing interpretes and compiles these assignments into a practical schedule for the participating vehicles.

\subsection{Hardware Implementation and Error Mitigation}

We executed the QAOA circuit on an IBM Quantum superconducting quantum processor. After obtaining the final parameterized ansatz circuit (defined by the optimized QAOA parameters from the classical tuning loop), we transpiled it to match the specific constraints of the quantum device. This transpilation pass involves rewriting and optimizing the circuit's gates to align with the hardware's native gate set and coupling map, minimizing qubit idle times and reducing gate count wherever possible. We employed the \emph{Sampler} primitive offered by IBM Quantum, along with hardware-aware error mitigation strategies. Each circuit execution included the following:
\begin{itemize}
    \item \textbf{Dynamical Decoupling}: We enabled a dynamical decoupling sequence with \texttt{XY4} that inserts idle pulses to reduce dephasing and other noise effects on qubits that remain idle while other qubits undergo gate operations.
    \item \textbf{Twirling}: We applied random gate twirling to symmetrize errors. This approach introduces random \emph{Clifford} operations before and after each gate block to average out coherent error effects, turning them into more uniform, easier-to-mitigate depolarizing noise.
    \item \textbf{Multiple Shots and Post-processing}: A sufficiently large number of measurement shots ranging from 10,000 to 16,000  were collected to extract statistically meaningful bitstring distributions. We then identified high-probability bitstrings as candidate solutions. 
\end{itemize}

\section{Quantum Results}
\label{sec:experimental_results}

In this section, we present the results obtained from both simulator and hardware experiments using QAOA for different problem sizes, namely for 2, 3, and 4 surfer–breaker pairs. The simulator results were obtained using the Qiskit Aer simulator, while the hardware results were gathered on IBM quantum hardware. 

\subsection{Simulator and Hardware Results}
Table~\ref{tab:results-compare} summarizes the simulator and hardware outcomes. For each problem instance, we list the number of surfer–breaker pairs, the corresponding number of qubits, the optimal QUBO cost value, and the most likely bitstring extracted from the measurement distribution.

\begin{table}[t]
\caption{Simulator vs.\ hardware: cost and top bitstrings.  
Classical optimum in bold; † = incorrect assignment.}
\label{tab:results-compare}
\centering
\begin{tabular}{ccccc}
\toprule
Pairs & Backend & Cost & Best bitstring & Bit-Sim \\ \midrule
2 & Sim. & 50 441.00 & \textbf{1001} & 1.00\\
  & HW   & 50 441.08 & \textbf{1001} & 1.00\\
\midrule
4 & Sim. & 113 876.29 & \textbf{1000001001000001} & 1.00\\
  & HW   & 113 879.80 & $\dagger1000001000010100$ & 0.75\\
\bottomrule
\end{tabular}
\end{table}
\footnote{Bit-sim refers to Bit-Similarity Index defined in Subsection \ref{ssec:bit-sim}.}

Table~\ref{tab:results-compare} also compares the results obtained from running the QAOA circuits on IBM quantum hardware versus the simulator results which are proven to be in agreement with the classical optima. The hardware experiments confirm that the optimal parameter sets found via the grid search are effective across a range of problem sizes. The only limitation appears to be quantum noise. As can be seen in Table~\ref{tab:results-compare}, the hardware did not reproduce the simulator result for the $16$ qubit problem with 4 surfers and 4 breakers.



\subsection{Decoding the QUBO Solution}
The QUBO model contains one binary variable \(x_{s,b}\) for every potential
pairing of wind-\textbf{surfer} \(s\in\{0,\dots,n-1\}\) with wind-\textbf{breaker}
\(b\in\{0,\dots,n-1\}\); hence there are \(n^{2}\) variables in total.
The solver returns a bit-string
\[
\boldsymbol{x}=(x_{0,0},\,x_{0,1},\dots,x_{0,n-1},\;
                x_{1,0},\dots,x_{n-1,n-1})\in\{0,1\}^{n^{2}}
\]
which are \emph{row-major} ordered .
Decoding the matrix into edge assignments is a linear-time post-processing stage that reshapes the length-$n^{2}$ QUBO bitstring into an $n\times n$ assignment matrix.  A single pass checks that every row and column contains exactly one~``1’’; otherwise the bitstring is flagged as infeasible (See Algorithm~\ref{alg:decode}).  The routine therefore runs in $\mathcal{O}(n^{2})$ time and $\mathcal{O}(n^{2})$ memory, negligible compared with solving the QUBO itself.

\begin{algorithm}[t]
\caption{\textsc{DecodeQUBOSolution}}
\label{alg:decode}
\begin{algorithmic}[1]
\Require Bitstring $x$ of length $n^{2}$ in **row-major** order  
          \hfill{\small(\(x_{1,1},\dots,x_{1,n},\,x_{2,1},\dots,x_{n,n}\))}
\Ensure  Matching $\mathcal{M}$ between surfers and breakers,  
         or \textbf{invalid} if constraints are violated
\State $n \gets \sqrt{|x|}$ \Comment{must be an integer}
\State Reshape $x$ into matrix $A \in \{0,1\}^{n \times n}$ \Comment{$A_{s,b}=1$ iff $s\mapsto b$}
\State $\mathcal{M} \gets \emptyset$
\For{$s \gets 0$ \textbf{to} $n-1$}                            \Comment{row uniqueness}
    \If{$\sum_{b=0}^{n-1} A_{s,b} \neq 1$}
        \State \Return \textbf{invalid}
    \EndIf
    \State $b^\star \gets \text{index such that } A_{s,b^\star}=1$
    \State $\mathcal{M} \gets \mathcal{M} \cup \{(s,b^\star)\}$
\EndFor
\For{$b \gets 0$ \textbf{to} $n-1$}                            \Comment{column uniqueness}
    \If{$\sum_{s=0}^{n-1} A_{s,b} \neq 1$}
        \State \Return \textbf{invalid}
    \EndIf
\EndFor
\State \Return $\mathcal{M}$
\end{algorithmic}
\end{algorithm}

\medskip
\noindent
As an example consider the following solution bit-string obtained from a hypothetical solver,
\(\mathtt{1000001001000001}\) which reshapes to
\[
\bordermatrix{~ & b=0 & 1 & 2 & 3\cr
 s=0 & 1&0&0&0\cr
      1 & 0&0&1&0\cr
      2 & 0&1&0&0\cr
      3 & 0&0&0&1\cr}\!.
\]
Reading off the `1's yields the matching
\(\{(0,0),(1,2),(2,1),(3,3)\}\) as we have in Figure ~\ref{fig:hungarian-demo}.


\subsection{The Bit-Similarity Metric and Runtimes}
\label{ssec:bit-sim}

To quantitatively assess the future potential of quantum optimization beyond the NISQ era, we explore the run times and bit-similarity of QAOA outputs without quantum error mitigation. We define bit-similarity index based on the normalized Hamming similarity between the target and output bitstrings. Given a target bitstring $x = (x_1, x_2, \dots, x_n)$ and an output bitstring $y = (y_1, y_2, \dots, y_n)$ of equal length $n$, the \emph{bitstring overlap similarity} $\mathcal{S}(x, y)$ is defined as:
\begin{equation}
\mathcal{S}(x, y) = \frac{1}{n} \sum_{i=1}^{n} \delta_{x_i, y_i},
\end{equation}
where $\delta_{x_i, y_i}$ is the Kronecker delta function, defined as:
\begin{equation}
\delta_{x_i, y_i} =
\begin{cases}
1, & \text{if } x_i = y_i, \\
0, & \text{otherwise}.
\end{cases}
\end{equation}

This metric quantifies the fraction of bit positions where the two bitstrings match. A score of $\mathcal{S}(x, y) = 1$ indicates a perfect match, while $\mathcal{S}(x, y) = 0$ implies no overlap at all. It provides insight into how closely the QAOA results approximate the optimal solution and has been used in prior QAOA benchmarking studies to evaluate solution quality in other combinatorial optimization problems \cite{zhou2020quantum} \cite{TransferabilityOO}. Whereas 2 surfer-breaker pairs and 3 surfer-breaker pairs achieve a perfect Bit-Similarity index on the Qiskit simulator and on the QPU, the 4 surfer-breaker example achieves only a score of 0.75 on the quantum hardware indicating the onset of noise-induced limitations.

\begin{figure}[H]
    \centering
    \fbox{\includegraphics[width=0.45\textwidth]{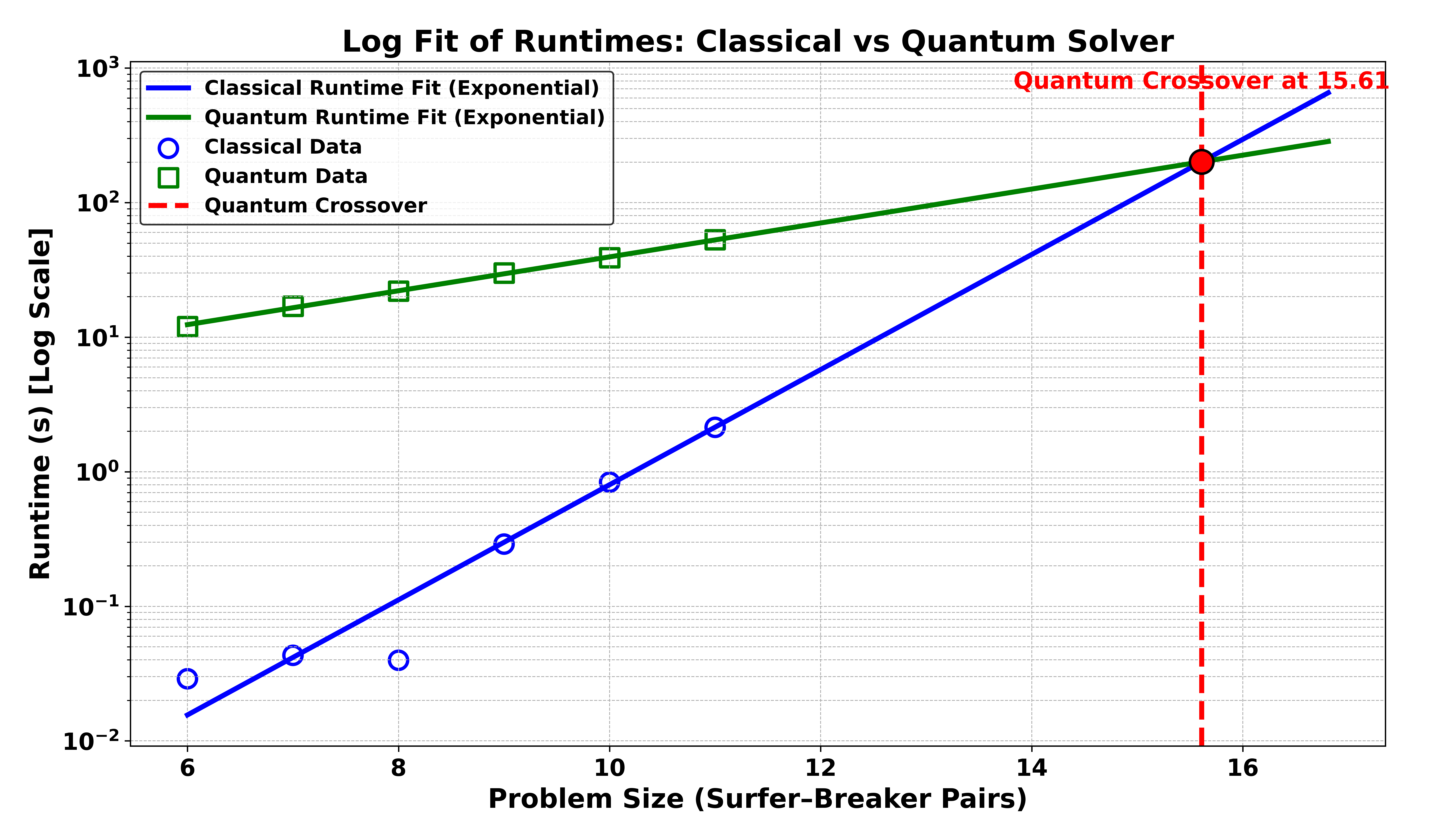 }}
    \caption{Runtime of the Gurobi based classical QUBO solver vs.\ Quantum Hardware time for a fixed depth (4) QAOA, as a function of problem size with a  log scale fit. The fit captures the rising trend in the classical computation time to be faster than the quantum counterpart. Both solvers are fitted with an exponential curve, adopting the worst case scenario for the quantum runtime. The marked intersection point indicates the \textbf{Quantum Runtime Advantage} threshold where the classical runtime exceeds the quantum runtime. However, about 256 qubits would be needed to encode a problem near the crossover point.}
    \label{fig:gurobi_fit}
\end{figure}

Table~\ref{tab:benchmark-results} presents the benchmark results for evaluating the runtime scaling and performance of both the classical and quantum heuristic approaches across varying problem sizes, defined by the number of surfer-breaker pairs ranging from 6 (36 qubits) to 11 (121 qubits). While the runtime scaling of the quantum hardware is encouraging, the bit-similarity index shows that hardware noise remains one of the most significant obstacles to quantum utility. Fig.~\ref{fig:gurobi_fit} provides additional indications of the future potentials of quantum optimization for this problem for larger problem sizes.

\begin{table}[ht]
\centering
\resizebox{\columnwidth}{!}{%
\begin{tabular}{ccccc}
\toprule
\textbf{Problem Size} & \textbf{Gurobi Time (s)} & \textbf{Quantum Time (s)} & \textbf{Hardware Time (s)} & \textbf{Bit-Similarity Index} \\
\midrule
6  & 0.02895 & 78.79166  & 12 & 0.53 \\
7  & 0.04324 & 46.54481  & 17 & 0.48 \\
8  & 0.03971 & 234.95370 & 22 & 0.48 \\
9  & 0.29004 & 130.14712 & 30 & 0.44 \\
10 & 0.83686 & 84.63909  & 39 & 0.47 \\
11 & 2.13769 & 144.98255 & 53 & 0.49 \\
\bottomrule
\end{tabular}%
}
\caption{Benchmark results for evaluating the runtime scaling and performance of both classical (Gurobi) and quantum (QAOA) approaches across varying problem sizes, defined by the number of surfer-breaker pairs ranging from 6 to 11.}
\label{tab:benchmark-results}
\end{table}

\paragraph*{Outlook}
A companion study \cite{Onah2025Trieste} extends this benchmarking effort across a much larger set of problems, various optimization engines, and several performance metrics.  



\section{Conclusion}
\label{sec:conclusion}

This paper has introduced ``Windbreaking-as-a-Service'' (WaaS) as a flexible, low-tech platooning concept and demonstrated how quantum optimization methods can form the computational backbone for its complex matching and assignment problems. We have systematically formulated the resulting vehicle matching and assignment problem as a Mixed-Integer Quadratic Programming (MIQP) task and presented its translation into a Quadratic Unconstrained Binary Optimization (QUBO) model with multi-segment routes, timing, class-based efficiency, and handover constraints. By translating WaaS into a QUBO model, we leverage classical methods for validation and employ QAOA for exploration of near-term quantum heuristics. Experimental results on both simulators and quantum hardware suggest that, while a definitive quantum speedup remains unproven, hybrid methods show promising solution quality and may yield real-world benefits faster than classical computers as hardware scales beyond 256 qubits. Future work will broaden the computational study of QUEST to multi-route highways and investigate its performance on large-scale shared-transportation optimization problems, moving closer to energy-efficient, low-emission mobility solutions.

\section{Acknowledgment}
We gratefully acknowledge Agneev Guin, Alejandro, Montanez-Barrera, Thorsten Grahs, and  Arne-Christian Voigt for fruitful feedback and discussion. 
\balance

\end{document}